# PaPy: Parallel and distributed data-processing pipelines in Python

Marcin Cieślik (mpc4p@virginia.edu) – *University of Virginia*, U.S.
Cameron Mura (cmura@virginia.edu) – *University of Virginia*, U.S.

**PaPy, which stands for parallel pipelines in Python, is a highly flexible framework that enables the construction of robust, scalable workflows for either generating or processing voluminous datasets. A workflow is created from user-written Python functions (nodes) connected by 'pipes' (edges) into a directed acyclic graph. These functions are arbitrarily definable, and can make use of any Python modules or external binaries. Given a user-defined topology and collection of input data, functions are composed into nested higher-order maps, which are transparently and robustly evaluated in parallel on a single computer or on remote hosts. Local and remote computational resources can be flexibly pooled and assigned to functional nodes, thereby allowing facile load-balancing and pipeline optimization to maximize computational throughput. Input items are processed by nodes in parallel, and traverse the graph in batches of adjustable size - a trade-off between lazy-evaluation, parallelism, and memory consumption. The processing of a single item can be parallelized in a scatter/gather scheme. The simplicity and flexibility of distributed workflows using PaPy bridges the gap between desktop -> grid, enabling this new computing paradigm to be leveraged in the processing of large scientific datasets.**

## Introduction

Computationally-intense fields ranging from astronomy to chemoinformatics to computational biology typically involve complex workflows of data production or aggregation, processing, and analysis. Several fundamentally different *forms* of data - sequence strings (text files), coordinates (and coordinate trajectories), images, interaction maps, microarray data, videos, arrays - may exist in distinct file formats, and are typically processed using available tools. Inputs/outputs are generally linked (if at all) *via* intermediary files in the context of some automated build software or scripts. The recently exponential growth of datasets generated by high-throughput scientific approaches (*e.g.* structural genomics [TeStYo09]) or high-performance parallel computing methods (*e.g.* molecular dynamics [KlLiDr09]) necessitates more flexible and scalable tools at the consumer end, enabling, for instance, the leveraging of multiple CPU cores and computational grids. However, using files to communicate and synchronize processes is generally inconvenient and inefficient, particularly if specialized scientific Python modules (*e.g.*, BioPython [CoAnCh09],

PyCogent [Knight07], Cinfony [OBHu08], MMTK [Hinsen00], Biskit [GrNiLe07]) are to be used.

Many computational tasks fundamentally consist of chained transformations of collections of data that are independent, and likely of variable type (strings, images, *etc.*). The scientific programmer is required to write transformation steps, connect them and - for large datasets to be feasible - parallelize the processing. Solutions to this problem generally can be divided into: *(i) Make*-like software build tools, *(ii)* workflow management systems (WMS), or *(iii)* grid engines and frontends. PaPy, which stands for parallel pipelines in Python, is a module for processing arbitrary streams of data (files, records, simulation frames, images, videos, *etc.*) *via* functions connected into directed graphs (flowcharts) like a WMS. It is not a parallel computing paradigm like *MapReduce* [DeGh08] or BSP [SkHiMc96], nor is it a dependency-handling build tool like Scons [Knight05]. Neither does it support declarative programming [Lloyd94]. In a nutshell, PaPy is a tool that makes it easy to structure procedural workflows into Python scripts. The tasks and data are composed into nested higher-order map functions, which are transparently and robustly evaluated in parallel on a single computer or remote hosts.

Workflow management solutions typically provide a means to connect standardized tasks *via* a structured, well-defined data format to construct a workflow. For transformations outside the default repertoire of the program, the user must program a custom task with inputs and outputs in some particular (WMS-specific) format. This, then, limits the general capability of a WMS in utilizing available codes to perform non-standard or computationally-demanding analyses. Examples of existing frameworks for constructing data-processing pipelines include Taverna (focused on webservices; run locally [OiAdFe04]), DAGMan (general; part of the Condor workload management system [ThTaLi05]) and Cyrille2 (focused on genomics; run on SGE clusters [Ham08]). A typical drawback of integrated WMS solutions such as the above is that, for tasks which are not in the standard repertoire of the program, the user has to either develop a custom task or revert to traditional scripting for parts of the pipeline; while such an approach offers an immediate solution, it is not easily sustainable, scalable, or adaptable, insofar as the *processing logic* becomes hardwired into these script-based workflows.

In PaPy, pipelines are constructed from Python functions with strict call semantics. Most general-purpose functions to support input/output, databases, inter-process communication (IPC), serialization, topology, and mathematics are already a part of PaPy. Domain-specific functions (*e.g.* parsing a specific file-format) must be user-provided, but have no limitations as to functional complexity, used libraries, called binaries or web-services, *etc*. Therefore, as a general pipeline construction tool, PaPy is intentionally lightweight, and is entirely agnostic of specific application domains. Our approach with PaPy is a highly modular workflow-engine, which neither enforces a particular data-exchange or restricted programming model, nor is tied to a single, specific application domain. This level of abstraction enables existing code-bases to be easily wrapped into a PaPy pipeline and benefit from its robustness to exceptions, logging, and parallelism.

## Architecture and design

PaPy is a Python module "papy" written to enable the logical design and deployment of efficient data-processing pipelines. Central design goals were to make the framework *(i)* natively parallel, *(ii)* flexible, *(iii)* robust, *(iv)* free of idiosyncrasies and dependencies, and *(v)* easily usable. Therefore, PaPy's modular, object-oriented architecture utilizes familiar concepts such as *map* constructs from functional programming, and directed acyclic graphs. Parallelism is achieved through the shared worker-pool model [Sunderam90]. The architecture of PaPy is remarkably simple, yet flexible. It consists of only four core component *classes* to enable construction of a data-processing pipeline. Each class provides an isolated subset of the functionality [Table1], which together includes facilities for arbitrary flow-chart topology, execution (serial, parallel, distributed), user function wrapping, and run-time interactions (*e.g.* logging). The pipeline is a way of expressing **what** (*functions*), **where** (*toplogy*) and **how** (*parallelism*) a collection of (potentially interdependent) calculations should be performed.

**Table 1: Components (*classes*) and their roles.**

| Component | Description and function |
|---|---|
| IMap[1] | Implements a process/thread pool. Evaluates multiple, nested map functions in parallel, using a mixture of threads or processes (locally) and, optionally, remote RPyC servers. |
| Piper Worker | Processing nodes of the pipeline created by wrapping user-defined functions; also, exception handling, logging, and scatter-gather functionality. |
| Dagger | Defines the data-flow and the pipeline in the form of a directed acyclic graph (DAG); allows one to add, remove, connect pipers, and validate topology. Coordinates the starting/stopping of IMaps. |
| Plumber | Interface to monitor and run a pipeline; provides methods to save/load pipelines, monitor state, save results. |

Pipelines (see Figure 1) are constructed by connecting functional units (`Piper` instances) by directed pipes, and are represented as a directed acyclic graph data structure (`Dagger` instance). The pipers correspond to nodes and the pipes to edges in a graph. The topological sort of this graph reflects the input/output dependencies of the pipers, and it is worth noting that any valid DAG is a valid PaPy pipeline topology (*e.g.*, pipers can have multiple incoming and outgoing pipes, and the pipeline can have multiple inputs and outputs). A pipeline input consists of an iterable collection of data items, *e.g.* a list. PaPy does not utilize a custom file format to store a pipeline; instead, pipelines are constructed and saved as executable Python code. The PaPy module can be arbitrarily used within a Python script, although some helpful and relevant conventions to construct a workflow script are described in the online documentation.

The functionality of a piper is defined by user-written functions, which are Python functions with strict call semantics. There are no limits as to what a function does, apart from the requirement that any modules it utilizes must be available on the remote execution hosts (if utilizing RPyC). A function can be used by multiple pipers, and multiple functions can be composed within a single piper. CPU-intensive tasks with little input data (*e.g.*, MD simulations, collision detection, graph matching) are preferred because of the high speed-up through parallel execution.

Within a PaPy pipeline, data are shared as Python objects; this is in contrast to workflow management solutions (*e.g.*, Taverna) that typically enforce a specific data exchange scheme. The user has the choice to use any or none of the structured data-exchange formats, provided the tools for using them are available for Python. Communicated Python objects need to be serializable, by default using the standard Pickle protocol.

Synchronization and data communication between pipers within a pipeline is achieved by virtue of queues and locked pipes. No outputs or intermediate results are implicitly stored, in contrast to usage of temporary files by *Make*-like software. Data can be saved anywhere within the pipeline by using pipers for data serialization (*e.g.* JSON) and archiving (*e.g.* file-based). PaPy maintains data integrity in the sense that an executing pipeline stopped by the user will have no pending (lost) results.

## Parallelism

Parallel execution is a major issue for workflows, particularly *(i)* those involving highly CPU-intensive methods like MD simulations or Monte Carlo sampling, or *(ii)* those dealing with large datasets (such as arise in astrophysics, genomics, *etc*.). PaPy provides support for two levels of parallelism, which adress both of these scenarios: (1) parallel processing of independent input data items, (2) multiple parallel jobs for a single input item. The first type of parallelism is achieved by creating parallel pipers - *i.e.* providing an IMap instance to the constructor. Pipers within a pipeline can share an IMap instance or have dedicated computational resources (Fig. 1). The mixing of serial and parallel pipers is supported; this flexibility permits intelligent pipeline load-balancing and optimization. Per-item parallel jobs are made possible by the *produce / spawn / consume* (Fig. 2) idiom within a workflow. This idiom consists of at least three pipers. The role of the first piper is to produce a list of $N$ subitems for each input item. Each of those subitems is processed by the next piper, which needs to be spawned $N$ times; finally, the $N$ results are consumed by the last piper, which returns a single result. Multiple spawning pipers are supported. The subitems are typically independent fragments of the input item or parameter sets. Per-item parallelism is similar to the *MapReduce* model of distributed computing, but is not restricted to handling only data structured as ($key$, $value$) pairs.

---

[1]Note that the IMap class is available as a separate Python module.

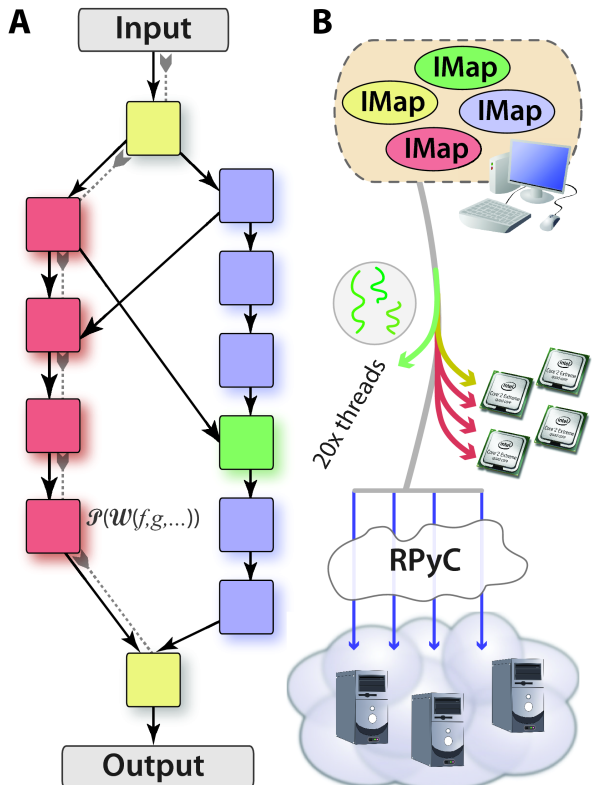

**Figure 1.** (**A**) PaPy pipeline and its (**B**) computational resources. The directed graph illustrates the Dagger object as a container of Piper objects (nodes), connected by pipes (black arrows; in the upstream / downstream sense) or, equivalently, dependency edges (gray arrows). Pipers are assigned to various compute resources as indicated by different colors. The sets of pipes connecting the two processing streams illustrate the flexible construction of workflows. Encapsulation and composition of user-written functions e.g., f, g into a Worker and Piper object is represented as P(W(f,g,...)). Resources used by the sample pipeline are shown in B. Each virtual resource is an IMap object, which utilizes a worker pool to evaluate the Worker on a data item. IMaps are shared by pipers and might share resources. The resources are: a local pool of 20 threads used by a single piper in the pipeline (green); four CPU-cores, of which at most three are used concurrently (red) and one dedicated to handle the input/output functions (yellow); and a pool of Python processes utilizing remote resources exposed by RPyC servers (blue cloud). Parallelism is achieved by pulling data through the pipeline in adjustable batches.

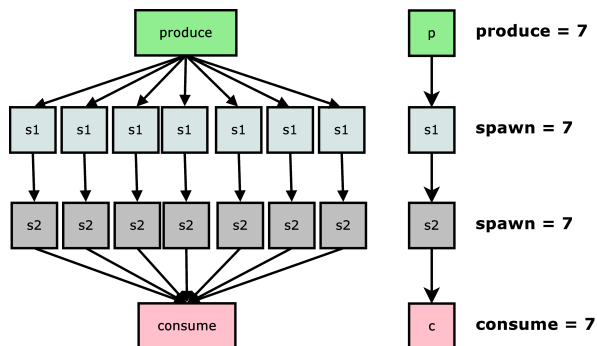

**Figure 2.** The produce / spawn / consume idiom allows for parallel processing of a single input item in addition to parallel processing of items (explanation in text).

The parallelism of an IMap instance is defined by the number of local and remote worker processes or threads, and the "stride" argument (Fig. 3), if it processes multiple tasks. The "stride" is the number of input items of task $N$ processed before task $N+1$ commences. Tasks are cycled until all input items have been processed. In a PaPy pipeline pipers can share a computational resource; they are different tasks of a single IMap instance. The "stride" can also be considered as the number of input items processed by pipers in consecutive rounds, with the order defined by a topological sort of the graph. Therefore, the data traverses the pipeline in batches of "stride" size. A larger "stride" means that potentially more temporary results will have to be held in memory, while a smaller value may result in idle CPUs, as a new task cannot start until the previous one finishes its "stride". This adjustable memory/parallelism trade-off allows PaPy pipelines to process data sets with temporary results too large to fit into memory (or to be stored as files), and to cope with highly variable execution times for input items (a common scenario on a heterogenous grid, and which would arise for certain types of tasks, such as replica-exchange MD simulations).

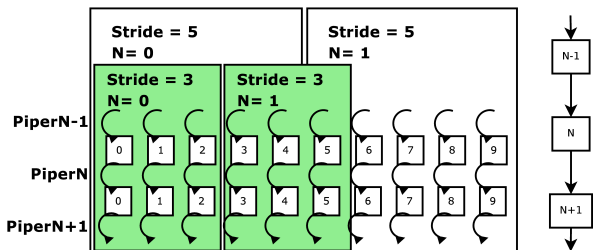

**Figure 3.** *The stride as a trade-off between memory consumption and parallelism of execution. Rectangular boxes represent graph traversal in batches. The pipers involved (N-1, N, N+2) are shown on the right (explanation in text).*

**Inter-process communication**

A major aspect - and often bottleneck - of parallel computing is inter-process communication (IPC; Fig. 4) [LiYa00]. In PaPy, IPC occurs between parallel pipers connected in a workflow. The communication process is two-stage and involves a manager process - *i.e*, the local Python interpreter used to start the workflow (Fig. 4). A coordinating process is necessary because the connected nodes might evaluate functions in processes with no parent/child relationship. If communication occurs between processes on different hosts, an additional step of IPC (involving a local and a remote RPyC process) is present. Inter-process communication involves data serialization (*i.e.* representation in a form which can be sent or stored), the actual data-transmission (*e.g.* over a network socket) and, finally, de-serialization on the recipient end. Because the local manager process is involved in serializing (de-serializing) data to (from) each parallel process, it can clearly limit pipeline performance if large amounts of data are to be communicated.

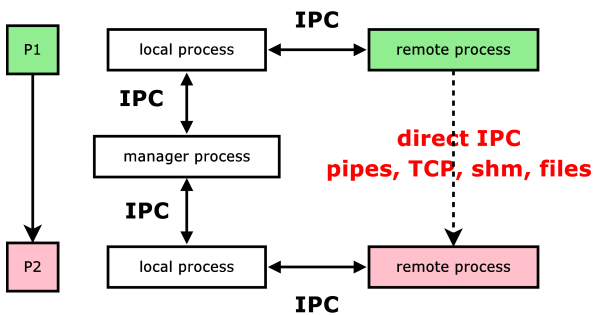

**Figure 4.** *Inter-process communication (IPC) between pipers (p1, p2). The dashed arrow illustrates possible direct IPC. Communication between the local and remote processes utilizes RPyC (explanation in text).*

PaPy provides functionality for direct communication of producer and consumer processes, thereby mostly eliminating the manager process from IPC and alleviating the bottleneck described above. Multiple serialization and transmission media are supported. In general, the producer makes data available (*e.g.* by serializing it and opening a network socket) and sends only information needed by the consumer end to locate the data (*e.g.* the host and port of the network socket) *via* the manager process. The consumer end receives this information and reads the data. Direct communication comes at the cost of losing platform-independence, as the operating system(s) have to properly support the chosen transmission medium (*e.g.* Unix pipes). Table 2 summarizes PaPy's currently available options.

**Table 2: Direct inter-process communication methods.**[2]

| Method | OS | Remarks |
|---|---|---|
| socket | all | Communication between hosts connected by a network. |
| pipe | UNIX-like | Communication between processes on a single host. |
| file | all | The storage location needs to be accessible by all processes - *e.g* over NFS or a SAMBA share. |
| shm | POSIX | Shared memory support is provided by the `posix_shm` library; it is an alternative to communication by pipes. |
| database | all | Serialized data can be stored as (key, value) pairs in a database. The keys are semi-random. Currently SQLite and MySQL are supported, as provided by mysql-python and sqlite3. |

Note that it is possible to avoid some IPC by logically grouping processing steps within a single piper. This is done by constructing a single piper instance from a worker instance created from a tuple of user-written functions, instead of constructing multiple piper instances from single function worker instances. A worker instance is a callable object passed to the constructor of the Piper class. Also, note that any linear, non-branching segment of a pipeline can be collapsed into a single piper. This has the performance advantage that no IPC occurs between functions within a single piper, as they are executed in the same process.

**Additional features and notes**

Workflow logging

PaPy provides support for detailed workflow logging and is robust to exceptions (errors) within user-written functions. These two features have been a major design goal. Robustness is achieved by embedding calls to user functions in a `try ... except` clause. If an exception is raised, it is caught and does not stop the execution of the workflow (rather, it is wrapped and passed as a placeholder). Subsequent pipers ignore and propagate such objects. Logging is supported via the `logging` module from the Python standard library. The `papy` and `IMap` packages emit logging statements at several levels of detail, *i.e.* DEBUG, INFO, ERROR; additionally, a function to easily setup and save or display logs is included. The log is written realtime, and can be used to monitor the execution of a workflow.

---

[2] Currently supported serialization algorithms: pickle, marshall, JSON

### Usage notes

A started parallel piper consumes a sequence of $N$ input items (where $N$ is defined by the "stride" argument), and produces a sequence of $N$ resultant items. Pipers are by default "ordered", meaning that an input item and its corresponding result item have the same index in both sequences. The order in which result items become available may differ from the order input items are submitted for parallel processing. In a pipeline, result items of an upstream piper are input items for a downstream piper. The downstream piper can process input items only as fast as result items are produced by the upstream piper. Thus, an inefficency arises if the upstream piper does not return an available result because it is out of order. This results in idle processes, and the problem can be addressed by using a "stride" larger then the number of processes, or by allowing the upstream piper to return results in the order they become available. The first solution results in higher memory consumption, while the second irreversibly abolishes the original order of input data.

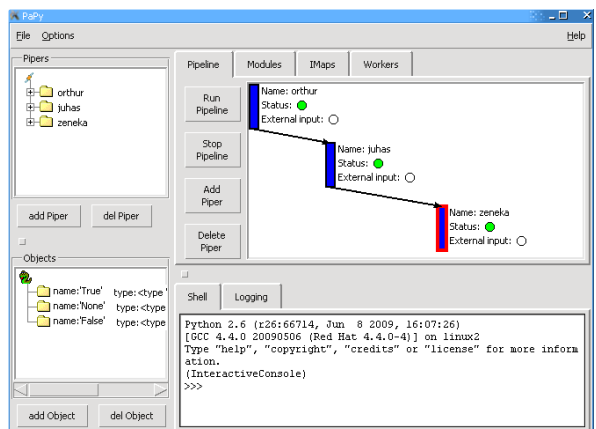

**Figure 5.** *A screenshot of the PaPy GUI written in Tkinter. Includes an interactive Python console and an intuitive canvas to construct workflows.*

### Graphical interface

As a Python package, PaPy's main purpose is to supply and expose an API for the abstraction of a parallel workflow. This has the advantage of flexibility (*e.g.* usage within other Python programs), but requires that the programmer learn the API. A graphical user interface (GUI) is currently being actively developed (Fig. 5). The motivation for this functionality is to allow a user to interactively construct, execute (*e.g.* pause execution), and monitor (*e.g.* view logs) a workflow. While custom functions will still have to be written in Python, the GUI liberates the user from knowing the specifics of the PaPy API; instead, the user explores the construction of PaPy workflows by connecting objects *via* navigation in the GUI.

### Workflow construction example

The following code listing illustrates steps in the construction of a distributed PaPy pipeline. The first of the two nodes evaluates a function (which simply determines the host on which it is run), and the second prints the result locally. The first piper is assigned to a virtual resource combining local and remote processes. The scripts take two command line arguments: a definition of the available remote hosts and a switch for using TCP sockets for direct inter-process communication between the pipers. The source code uses the `imports` decorator. This construct allows import statements to be attached to the code of a function. As noted earlier, the imported modules must be available on all hosts on which this function is run.

The pipeline is started, for example, *via*:

```
$ python pipeline.py \
        --workers=HOST1:PORT1#2,HOST2:PORT1#4
```

which uses 2 processes on `HOST1` and 4 on `HOST2`, and all locally-available CPUs. Remote hosts can be started (assuming appropriate firewall settings) by:

```
$ python RPYC_PATH/servers/classic_server.py \
        -m forking -p PORT
```

This starts a RPyC server listening on the specified PORT, which forks whenever a client connects. A forking server is capable of utilizing multiple CPU cores. The following example (in expanded form) is provided as part of PaPy's online documentation.:

```
#!/usr/bin/env python
# Part 0: import the PaPy infrastructure.
# papy and IMap are separate modules
from papy import Plumber, Piper, Worker
from IMap import IMap, imports
from papy import workers

# Part 1: Define user functions
@imports(['socket', 'os', 'threading'])
def where(inbox):
    result = "input: %s, host:%s, parent %s, \
                    process:%s, thread:%s" % \
    (inbox[0], \
    socket.gethostname(), \
    # the host name as reported by the OS
    os.getppid(), \ # get parent process id
    os.getpid(), \  # get process id
    threading._get_ident())
    # unique python thread identifier
    return result

# Part 2: Define the topology
def pipeline(remote, use_tcp):
    # this creates a IMap instance which uses
    #'remote' hosts.
    imap_ = IMap(worker_num=0, worker_remote=remote)
    # this defines the communication protocol i.e.
    # it creates worker instances with or without
    # explicit load_item functions.
    if not use_tcp:
        w_where = Worker(where)
        w_print = Worker(workers.io.print_)
    else:
        w_where = Worker((where, workers.io.dump_item), \
                         kwargs=({}, {'type':'tcp'}))
        w_print = Worker((workers.io.load_item, \
                          workers.io.print_))
    # the instances are combined into a piper instance
    p_where = Piper(w_where, parallel=imap_)
    p_print = Piper(w_print, debug=True)
    # piper instances are assembled into a workflow
    # (nodes of the graph)
    pipes = Plumber()
    pipes.add_pipe((p_where, p_print))
    return pipes

# Part 3: execute the pipeline
if __name__ == '__main__':
    # get command-line arguments using getopt
    # following part of the code is not PaPy specific
    # and has the purpose of interpreting commandline
    # arguments.
    import sys
    from getopt import getopt
    args = dict(getopt(sys.argv[1:], '', ['use_tcp=', \
                                          'workers='])[0])
    # parse arguments
    use_tcp = eval(args['--use_tcp']) # bool
    remote = args['--workers']
    remote = worker_remote.split(',')
    remote = [hn.split('#') for hn in remote]
    remote = [(h, int(n)) for h, n in remote]
    # create pipeline (see comments in function)
    pipes = pipeline(remote, use_tcp)
    # execution
    # the input to the function is a list of 100
    # integers.
    pipes.start([range(100)])
    # this starts the pipeline execution
    pipes.run()
    # wait until all input items are processed
    pipes.wait()
    # pause and stop (a running pipeline cannot
    # be stopped)
    pipes.pause()
    pipes.stop()
    # print execution statistics
    print pipes.stats
```

## Discussion and conclusions

In the context of PaPy, the factors dictating the computational efficiency of a user's pipeline are the nature of the individual functions (nodes, pipers), and the nature of the data linkages between the constituent nodes in the graph (edges, pipes). Although distributed and parallel computing methods are becoming ubiquitous in many scientific domains (*e.g.*, biologically meaningful usec-scale MD simulations [KlLiDrSh09]), data post-processing and analysis are not keeping pace, and will become only increasingly difficult on desktop workstations.

It is expected that the intrinsic flexibility underlying PaPy's design, and its easy resource distribution, could make it a useful component in the scientist's data-reduction toolkit. It should be noted that some data-generation workflows might also be expressible as pipelines. For instance, parallel tempering / replica-exchange MD [EaDe05] and multiple-walker metadynamics [Raiteri06] are examples of intrinsically parallelizable algorithms for exploration and reconstruction of free energy surfaces of sufficient granularity. In those computational contexts, PaPy could be used to orchestrate data *generation* as well as data aggregation / reduction / analysis.

In conclusion, we have designed and implemented PaPy, a workflow-engine for the Python programming language. PaPy's features and capabilities include: (1) construction of arbitrarily complex pipelines; (2) flexible tuning of local and remote parallelism; (3) specification of shared local and remote resources; (4) versatile handling of inter-process communication; and (5) an adjustable laziness/parallelism/memory trade-off. In terms of usability and other strengths, we note that PaPy exhibits (1) robustness to exceptions; (2) graceful support for time-outs; (3) real-time logging functionality; (4) cross-platform interoperability; (5) extensive testing and documentation (a 60+ page manual); and (6) a simple, object-oriented API accompanied by a preliminary version of a GUI.

## Availability

PaPy is distributed as an open-source, platform-independent Python (CPython 2.6) module at http://muralab.org/PaPy, where extensive documentation also can be found. It is easily installed *via* the Python Package Index (PyPI) at http://pypi.python.org/pypi/papy/ using `setuptools` by `easy_install papy`.

## Acknowledgements

We thank the University of Virginia for start-up funds in support of this research.